\documentstyle[aps,pre,eqsecnum,preprint]{revtex}

\begin{document}
\bibliographystyle{unsrt}
\textwidth 800pt

\large
\begin{center}
\underline{Tangent bifurcation of band edge
plane waves,} \\ 
\underline{dynamical symmetry breaking and vibrational
localization} 
\vspace{2cm}\\ \large S. \vspace{0.5cm}Flach$^*$ \\
\normalsize
Max-Planck-Institut f\"ur Physik Komplexer Systeme \\
Bayreuther Str.40 H.16, D-01187 Dresden, Germany \vspace{1cm} 
\\
\end{center}
\normalsize
ABSTRACT \\
We study tangent bifurcation of band edge plane waves
in nonlinear Hamiltonian lattices. The lattice is translationally
invariant. We argue for the breaking of permutational symmetry 
by the new bifurcated periodic orbits. The case of two 
coupled oscillators is considered as an example for the
perturbation analysis, where the symmetry breaking can be
traced using Poincare maps. Next we consider a lattice
and derive the dependence of the bifurcation energy 
on the parameters of the Hamiltonian function in the limit
of large system sizes. A necessary condition for the 
occurence of the bifurcation is the repelling of the
band edge plane wave from the linear spectrum with
increasing energy. We conclude that the bifurcated orbits
will consequently exponentially localize in the configurational
space. 
\vspace{0.5cm}
\newline
PACS number(s): 03.20.+i ; 63.20.Pw ; 63.20.Ry 
\newline
\newline
{\sl Physica D, accepted for publication} 
\newline
Date: 09/25/95
\\
\\
\\
$^*$ email: flach@idefix.mpipks-dresden.mpg.de

\newpage

\section{Introduction}

An increasing number of publications deal with the phenomenon
of discrete breathers in nonlinear translationally invariant
Hamiltonian lattices (for a history on the subject see
\cite{tks88},\cite{jbp90},\cite{cp90},\cite{st92},\cite{bks93},
\cite{fw2}, \cite{fw6},\cite{fw8} and references therein). 
Discrete breathers are time-periodic solutions of the equations
of motion. These solutions are exponentially localized in the
configurational space (see \cite{th91} and in detail \cite{fw7}). 
One can view discrete breathers
as analoga of the breather solutions of the sine-Gordon (sG)
partial differential equation \cite{cp90}. Breathers of the sG equation are
structurally unstable (nongeneric) against perturbations of the Hamiltonian
density of the sG field \cite{jd93},\cite{bb94}. 
In contrast discrete breathers appear to
be structurally stable (generic) solutions, 
with the seeming neccessary ingridients
being the nonlinearity and the discreteness (i.e. the existence of
a finite upper bound of the linear spectrum) \cite{fw9}. One could think that
the sG PDE is the 'bottleneck' which gives contact to the
discrete breathers by simply discretising the spatial differential
operator. This thought is most probably wrong, because in order
to maintain structural stability of discrete breathers one has
to avoid resonances of multiples of the breather's frequency with
the linear spectrum \cite{fw7}. 
Thus discrete breathers appear to be rather unrelated
to their counterpart in the sG case.

Proofs of existence of discrete breathers have been given so far
for i) arrays of weakly coupled anharmonic oscillators by use
of analytical continuation of periodic orbits \cite{ma94} and for ii)
chains with homogeneous interaction potentials by use of
discrete map analysis of the Fourier coefficients (with respect
to time) \cite{fw9}. Structural stability of discrete breathers has
been shown for one-dimensional lattices by use of discrete
map analysis of Fourier coefficients \cite{fw9}.

It turns out to be rather complicated to give a proof of existence
for discrete breathers for an arbitrarily choosen Hamiltonian function.
Still for applications or numerical studies it is important to
know whether a given system can posess discrete breathers.
From the Fourier coefficient analysis in \cite{fw7} we can predict,
that if the frequency of the breather comes close to the linear 
spectrum, then the exponent of the spatial decay becomes very small.
If the energy of the breather solution becomes small in the same
limit, then the breather solution comes close to the solution
of a band edge plane wave (here band edge plane wave stands
for a plane wave with a wave vector of the band edge of the
band of allowed frequencies of the linearized equations of motion). 
Consequently one could expect
that discrete breathers appear due to tangent bifurcations
of band edge plane waves if one increases the energy $E_{ZB}$ of
the plane wave starting from $E_{ZB}=0$. Indeed the birth of
discrete breather solutions as bifurcating periodic orbits due
to the tangent bifurcation of band edge waves
has been observed numerically for one-dimensional Fermi-Pasta-Ulam (FPU)
lattices \cite{sp94} with the number $N$ of lattice sites 
ranging between $N=40$ and $N=100$. Note that tangent bifurcation implies
that a pair of Floquet multipliers of the linear stability analysis of
a periodic orbit collide at +1 on the unit circle.

The bifurcation problem of band edge plane waves has been a problem
studied independently of the discrete breather phenomenon 
\cite{bb83},\cite{yak93}.
Budinsky and Bountis \cite{bb83} have approached the problem using Floquet
theory. They have then analyzed the eigenvalue problem for large
$N$. However there were discrepancies between the results in \cite{bb83}
and the study by Sandusky and Page \cite{sp94}. Also in both cases only
Fermi-Pasta-Ulam (FPU) chains were considered. Further no analytical
results about the properties of the new bifurcated orbits were given. 
Consequently the goal of the present paper is i) to derive exact
conditions for the tangent bifurcation of band edge plane waves
for large $N$, ii) to consider both FPU and nonlinear Klein-Gordon (KG)
systems, and 
iii) to show that the bifurcated orbits will break the permutational
symmetry of the system and exponentially localize in the configurational 
space.

In section II we analyze the bifurcational problem for a system
of two coupled oscillators ($N=2$). With the help of Poincare maps
we will show that the derived formula apply. In section III the 
general lattice with $N$ degrees of freedom is introduced, and the
periodic orbits of the band edge plane waves are obtained.
The results of the bifurcational analysis are presented in section IIIA,B.
In section IIIC we show that the bifurcated orbits break the permutational
symmetry. We discuss our results in section IV.

\section{The case of two coupled oscillators}

Let us consider the following Hamilton function
\begin{equation}
H=\frac{1}{2}\dot{X}_1^2 + \frac{1}{2}\dot{X}_2^2
+ V(X_1) + V(X_2) + \Phi(X_1 - X_2) \;\;\;, \label{1}
\end{equation}
which describes the dynamics of two coupled oscillators
using the equations of motion
\[
\ddot{X}_{1,2} = - \frac{\partial H}{\partial X_{1,2}} \;\;.
\]
The potential functions from (\ref{1}) can be expanded in the form
\begin{eqnarray}
V(z) = \sum_{\mu = 2}^{\infty} \frac{1}{\mu}v_{\mu}z^{\mu}\;\;, \label{2-1} \\
\Phi(z) = \sum_{\mu = 2}^{\infty} \frac{1}{\mu}\phi_{\mu}z^{\mu}\;\;. 
\label{2-2}   
\end{eqnarray}
Note that we have to demand $v_2 \geq 0$ and $\phi_2 > 0$ in order
to ensure the potential energy in (\ref{1}) being a positive definite
quadratic form in the limit of small energies.

\subsection{The normal mode periodic orbits}

For small energies we can neglect the anharmonic terms $\mu > 2$
in (\ref{2-1}),(\ref{2-2}). In that case it is appropriate to 
transform the original variables into normal coordinates
\begin{equation}
Y_1 = \frac{1}{2}(X_1 + X_2) \;\;,\;\;Y_2=\frac{1}{2}(X_1 - X_2)
\;\;. \label{3}        
\end{equation}
Using the relations given above we obtain the equations of motion
for the normal modes including all anharmonic terms of the potential
functions (\ref{2-1}),(\ref{2-2}):
\begin{eqnarray}
\ddot{Y}_1 = - \sum_{\mu=2}^{\infty}
v_{\mu} \sum_{\nu=0,2,...}^{(\nu \leq (\mu - 1))}
\left(
\begin{array}{c} \mu - 1 \\ \nu \end{array} \right)
Y_1^{\mu - 1 - \nu} Y_2^{\nu}\;\;, \label{4-1} \\
\ddot{Y}_2 = - \sum_{\mu=2}^{\infty}
v_{\mu} \sum_{\nu = 1,3,...}^{(\nu \leq (\mu - 1))}
\left(
\begin{array}{c} \mu - 1 \\ \nu \end{array} \right)
Y_1^{\mu - 1 - \nu}Y_2^{\nu} - \sum_{\mu = 2}^{\infty}
\phi_{\mu} (2Y_2)^{\mu - 1}\;\;. \label{4-2}      
\end{eqnarray}
Note that the sums over $\nu$ in these equations run over
even or odd values of $\nu$ respectively because cancellation
of terms due to (\ref{3}).

We can now solve (\ref{4-1},\ref{4-2}) immediately for the two
periodic orbits, which correspond to the standard harmonic
normal mode solutions in the limit of small energies:
\begin{eqnarray}
{\rm I}: \;\; Y_2=0\;\;,\;\;\ddot{Y}_1 = -\sum_{\mu=2}^{\infty}
v_{\mu}Y_1^{\mu-1}\;\;, \label{5-1} \\
{\rm II}: \;\; Y_1=0 \;\;,\;\; \ddot{Y}_2 = -\sum_{\mu=2}^{\infty}      
\tilde{v}_{\mu} Y_2^{\mu-1}\;\;,\;\;v_{2m+1}=0\;\;.\label{5-2}       
\end{eqnarray}
Here the renormalized constants $\tilde{v}_{\mu}$ for case II in
(\ref{5-2}) are given by
\begin{equation}
\tilde{v}_{\mu} = v_{\mu} + 2^{\mu - 1}\phi_{\mu}\;\;. \label{5-3}    
\end{equation}
Note that we had to require $v_{2m+1}=0$ in case II (\ref{5-2}),
since in the case of nonvanishing odd terms in the potential $V(z)$
the right hand side of (\ref{4-1}) has terms containing the
variable $Y_2$ only, and the solution would be characterized by
nonzero $Y_1$. 

The solutions of the differential equations (\ref{5-1}),(\ref{5-2}) 
are given in terms
of elliptic functions. Here we will be interested in the limit
of small energies (amplitudes) and apply standard perturbation
theory (\cite{ahn93} Chapter 7). For case I (equation (\ref{5-1}))
we obtain
\begin{eqnarray}
Y_1 = \sum_{k=-\infty}^{+\infty}A_k {\rm e}^{ik\omega t}\;\;,\;\;
A_k = A_{-k} \;\;, \label{6-1} \\
\omega^2 = v_2 +\left(\frac{3v_4}{2v_2} - \frac{5v_3^2}{3v_2^2}\right)E_{1}
\;\;, \label{6-2}\\
A_0 = -\frac{v_3}{v_2^2}E_1\;\;,\;\; A_1^2 =\frac{1}{2v_2}E_1 \;\;
, \;\; A_2 = \frac{v_3}{6v_2^2}E_1\;\;, \label{6-3} \\
E_1=\frac{1}{2}\dot{Y}_1^2 +V(Y_1) \;\;. \label{6-4}        
\end{eqnarray}
For case II (equation (\ref{5-2})) we obtain in a similar way
\begin{eqnarray}
Y_2 = \sum_{k=-\infty}^{+\infty}A_k {\rm e}^{ik\omega t}\;\;,\;\;
A_k = A_{-k} \;\;, \label{7-1} \\
\omega^2 = \tilde{v}_2 +\left(\frac{3\tilde{v}_4}{2\tilde{v}_2} 
- \frac{5\tilde{v}_3^2}{3\tilde{v}_2^2}\right)E_{2}
\;\;, \label{7-2}\\
A_0 = -\frac{\tilde{v}_3}{\tilde{v}_2^2}E_2\;\;,\;\; 
A_1^2 =-\frac{1}{2\tilde{v}_2}E_2 \;\;
, \;\; A_2 = \frac{\tilde{v}_3}{6\tilde{v}_2^2}E_2\;\;, \label{7-3} \\
E_2=\frac{1}{2}\dot{Y}_2^2 +\tilde{V}(Y_2)\;\;, \;\;
\tilde{V}(z) = \sum_{\mu=2}^{\infty}\frac{1}{\mu}\tilde{v}_{\mu}
z^{\mu}\;\;. \label{7-4}
\end{eqnarray}
All terms not present in (\ref{6-1})-(\ref{7-4}) are of order $E_{1,2}^2$
or higher and can be neglected, as we will see in the following.

\subsection{Tangent bifurcation for periodic orbit I}

Let us consider a small perturbation $\delta_{1,2}$ 
of the periodic orbit I
\[
Y_{1,2} \rightarrow Y_{1,2} + \delta_{1,2}\;\;.
\]
If we linearize the resulting equations
of motion for the perturbation we obtain
\begin{eqnarray}
\ddot{\delta}_1 = - \sum_{\mu = 2}^{\infty} (\mu-1)v_{\mu}Y_1^{\mu-2}
\delta_1 \;\;, \label{8-1} \\
\ddot{\delta}_2 = - \sum_{\mu=2}^{\infty}(\mu - 1)v_{\mu} Y_1^{\mu-2}
\delta_2 -2\phi_2 \delta_2 \;\;. \label{8-2}      
\end{eqnarray}
Note, that the variable $Y_1$ in (\ref{8-1}),(\ref{8-2}) is the time-periodic
solution of (\ref{5-1}).
As we see, the perturbations do not couple with each other. Equation
(\ref{8-1}) describes the continuation of the periodic orbit 
(\ref{5-1}) along itself (shift of time origin) or along the
one-parameter family (change of energy $E_1$ or frequency $\omega$).
The associated pair of Floquet multipliers is obviously located
at $+1$ on the unit circle \cite{ahn93}.

Equation (\ref{8-2}) then describes the nontrivial perturbations
(the associated phase space is two-dimensional, and can be visualized
with the help of Poincare maps). Let us rewrite (\ref{8-2}):
\begin{eqnarray}
\ddot{\delta}_2 = -\left[ (v_2 + 2\phi_2) + f(t) \right] \delta_2
\;\;, \label{9-1} \\
f(t) = \sum_{\mu=3}^{\infty} (\mu - 1) v_{\mu} Y_1^{\mu - 2}\;\;. 
\label{9-2}       
\end{eqnarray}
This type of equation is called Hill's equation \cite{mw79}, since the
function f(t) is periodic in time (with the period of $Y_1$):
\[
f(t) = f(t + \frac{2 \pi}{\omega})\;\;.
\]
We are interested in tangent bifurcations which appear when 
\begin{equation}
\frac{\sqrt{v_2 +2 \phi_2}}{\omega} \approx 1 \;\;.\label{9-3} 
\end{equation}
The function $f(t)$ in (\ref{9-2}) can be expanded in a Fourier series:
\begin{equation}
f(t) = \sum_{k=-\infty}^{+\infty} F_k {\rm e}^{ik\omega t} \;\;,\;\;
F(k) = F_{-k} \;\;. \label{10}    
\end{equation}
If the function $f(t)$ becomes infinitesimally small, then
according to the Floquet theory \cite{mw79} parametric resonance can appear
for the cases
\begin{equation}
\frac{\sqrt{v_2 + 2\phi_2}}{k\omega}= \frac{k'}{2}\;\;,\;\;k'=1,2,3,... \;
\;. \label{11}   
\end{equation}
With (\ref{9-3}) the relevant pairs $(k,k')$ become
$(1,2)\;\;,\;\;(2,1)$.    
In the limit of small values of $E_1$ we can then retain in (\ref{9-1})
the significant terms only:
\begin{eqnarray}
\ddot{\delta}_2 = -\left[ \Omega^2 + 2F_1{\rm cos}(\omega t)
+ 2F_2 {\rm cos}(2\omega t)\right] \delta_2 \;\;, \label{12-1} \\
\Omega^2 = v_2 + 2\phi_2 + F_0 \;\;.\label{12-2}   
\end{eqnarray}
Using relations (\ref{6-2}-\ref{6-4}) we obtain for small $E_1$
\begin{eqnarray}
F_0 = \left( -2\frac{v_3^2}{v_2^2} + 3 \frac{v_4}{v_2} \right) E_1
\;\;, \label{13-1} \\
F_1 = \sqrt{2}\frac{v_3}{\sqrt{v_2}} \sqrt{E_1}\;\;, \label{13-2} \\
F_2 = \left( \frac{1}{3}\frac{v_3^2}{v_2^2} + \frac{3}{2}\frac{v_4}{v_2}
\right) E_1 \;\;. \label{13-3}       
\end{eqnarray}
A tangent bifurcation will appear if the solution for the
perturbation $\delta_2(t)$ becomes periodic with the period
of the function $f(t)$ (since at the bifurcation the associated pair of 
Floquet multipliers has to merge at $+1$ on the unit circle):
\[
\delta_2(t) = \delta_2(t+\frac{2\pi}{\omega})\;\;.
\]
For given potential functions (\ref{2-1}),(\ref{2-2}) this can happen
only at certain energies $E_1^c$.
Applying the method of strained parameters \cite{ahn93} to our problem
we obtain two solutions
\begin{eqnarray}
{\rm i):} \;\; \phi_2 = 0\;\;,\;\; E_1^c \;\; {\rm arbitrary}\;\;,
\label{14-1} \\
{\rm ii):}\;\; E_1^c = \frac{6v_2^2}{10v_3^2 - 9v_2v_4}\phi_2 
\;\;. \label{14-2}      
\end{eqnarray}
Since we have to require positive values for $E_1$ and $\phi_2$
tangent bifurcation can take place only if
\begin{equation}
\frac{v_4}{v_2} \leq \frac{10}{9}\frac{v_3^2}{v_2^2}\label{15}
\end{equation}
is fulfilled. Remarkably this condition (\ref{15}) is equivalent
to requiring the frequency of the periodic orbit I being a
decreasing function of $E_1$ (cf. (\ref{6-2})). Note that we
have to exclude the nongeneric
case of $v_3=v_4=0$ (however we have no restrictions on $\phi_{3,4,...}$
since the anharmonic interaction terms simply do not contribute
to the linearized equations (\ref{8-1}),(\ref{8-2})).
The two solutions (\ref{14-1})-(\ref{14-2}) define two lines in the
space $\{ E_1,\phi_2\}$. The area between
the two lines corresponds to the parameter cases when the periodic
orbit is hyperbolic due to tangent bifurcation. Note that the
requirement of small energies implies also small values of $\phi_2$.
For large values of $E_1$ and $\phi_2$ higher order corrections will
appear. 

\subsection{Tangent bifurcation for periodic orbit II}

In analogy to the previous chapter we derive the linearized
equations for the perturbation of the periodic orbit II:
\begin{eqnarray}
\ddot{\delta}_1 = - \sum_{\mu=2}^{\infty}(\mu-1)v_{\mu}
Y_2^{\mu-2}\delta_1 \;\;,\label{16-1} \\
\ddot{\delta}_2=-\sum_{\mu=2}^{\infty}(\mu-1)\tilde{v}_{\mu}
Y_2^{\mu-2}\delta_2\;\;. \label{16-2}     
\end{eqnarray}
Note that we had to require $v_{2m+1}=0$. Again the perturbations
do not couple with each other. Equation (\ref{16-2}) describes
continuation of the periodic orbit II.

Equation (\ref{16-1}) gives the nontrivial perturbations. It is again
a Hill's equation. In the limit of small energies $E_2$
we obtain in analogy to (\ref{9-1})-(\ref{13-3})
\begin{eqnarray}
\ddot{\delta}_1 = -\left[ \Omega_2 + 2F_2 {\rm cos}(2\omega t) \right]
\delta_1 \;\;,\label{17-1} \\
\Omega^2 = v_2 + F_0 \;\;, \label{17-2} \\
F_0 = 3\frac{v_4}{\tilde{v}_2}E_2 \;\;, \label{17-3} \\
F_2 = \frac{3}{2}\frac{v_4}{\tilde{v}_2}E_2\;\;. \label{17-4}       
\end{eqnarray}
Requiring periodicity $\delta_1(t)=\delta_1(t+2\pi/\omega)$
and applying the method of strained parameters 
we find the two solutions for a tangent bifurcation
\begin{eqnarray}
{\rm i):}\;\; E_2^c = \frac{6 v_2^2}{80\phi_3^2 - 36v_2 \phi_4}\phi_2
\;\;, \label{18-1} \\
{\rm ii):}\;\;E_2^c = \frac{6 v_2^2}{9v_2v_4 + 80\phi_3^2 -36v_2\phi_4}
\phi_2 \;\;. \label{18-2}   
\end{eqnarray}
Again these two solutions define two lines in the space
$\{ E_2,\phi_2 \}$. The area between the lines corresponds
to the parameter cases when the periodic orbit II is hyperbolic
due to tangent bifurcation.

\subsection{Symmetry breaking}

The scenaria of the tangent bifurcations of periodic orbits
I and II from the last two chapters can be studied numerically
by use of Poincare maps. For that we always fix the total energy
of the system, and plot the pair $\{ X_1,\dot{X}_1 \}$
if the conditions $X_2=0$, $\dot{X}_2 > 0$ are fulfilled.
First we use this circumstance for checking the validity of our
results. We find excellent agreement. More precisely we test our
analytical calculations by evaluating numerically the Hill's eigenvalue
problem (\ref{8-2}) and (\ref{16-1}) for the two orbits. We find
the tangent bifurcation, i.e. the merging of the two Floquet
multipliers at +1 exactly for the predicted parameter combinations.  

An example for the bifurcation of periodic orbit I is
shown in Fig. 1 for the parameter cases $v_2=1$, $v_3=3$, $v_4=1$,
$\phi_2=0.01$, $\phi_3=1$, $\phi_4=1$. According to  
(\ref{14-2}) the bifurcation energy is given by $E_1^c=7.407\cdot 10^{-4}$.
In Fig.1a we show the Poincare map of the surrounding of the
periodic orbit I for $E_1=7\cdot 10^{-4}$ 
and in Fig.1b for $E_1=7.7\cdot 10^{-4}$
(note that we did not show Poincare maps too close to the bifurcation
point only because of the computational time which increases to infinity
at the bifurcation point because the winding number comes very close
to unity).
Clearly in Fig.1a the orbit I is still of elliptic character, whereas
in Fig.1b the bifurcation already occured, giving rise to the
birth of two new periodic orbits labelled by $A$ and $B$ in the 
plot respectively. These bifurcated orbits are separated by a
separatrix which contains the old periodic orbit I.

The bifurcation of periodic orbit II is demonstrated in Fig.2 for
the parameter cases $v_2=1$, $v_3=0$, $v_4=1$,
$\phi_2=0.001$, $\phi_3=1$, $\phi_4=0.1$. 
According to (\ref{18-1}),(\ref{18-2})
the bifurcation occurs for $E_2^c=7.0258\cdot 10^{-5}$. 
We plot the imaginary part of the Floquet multiplier evaluated
from equation (\ref{16-1}) as a function of the energy $E_2$.
The obtained value for the bifurcation energy is $E_2^c=7.042\cdot 10^{-5}$,
which gives a deviation of $0.2\%$ of the value from our perturbation
approach.  

The considered system of two coupled oscillators exhibits permutational
symmetry $H(X_1,X_2)=H(X_2,X_1)$ if we require $\phi_{2m+1}=0$. 
Still the tangent bifurcation of orbits I and II occur.
If we define the permutational operator $\hat{P}$ as
\[
\hat{P}g(X_1,X_2,\dot{X}_1,\dot{X}_2)=g(X_2,X_1,\dot{X}_2,\dot{X}_1)
\]
then the normal coordinates are transformed as
\begin{equation}
\hat{P}Y_1=Y_1\;\;,\;\;\hat{P}Y_2=-Y_2\;\;.\label{19}      
\end{equation}
Let us consider the bifurcation of orbit I. This orbit corresponds to
a closed loop in the phase space of the system. Its position
is restricted to the subspace $\{Y_2=\dot{Y}_2=0\}$. The trajectory of 
orbit I evolves along the loop parametrically with time. If a 
tangent bifurcation occurs, then the bifurcated orbits will
correspond to slightly deformed loops of the original one. These
deformed loops will move out of the subspace $\{Y_2=\dot{Y}_2=0\}$
as we have shown. 
Let us assume that the deformed loops are invariant under
permutation.
Applying $\hat{P}$ to such a loop is equivalent to a reflection
at the subspace$\{ Y_2=0,\dot{Y}_2=0\}$. It is clear that it is 
impossible to construct such a deformed loop without crossing
the subspace $\{ Y_2=0,\dot{Y}_2=0\}$. But it is forbidden to
cross this subspace, since any point from it belongs to a
periodic orbit I (parametrized by its energy). 
Thus we conclude that the deformed loops can not be invariant
under permutation. So the bifurcated orbits break permutational symmetry.
Then there have to exist at least two bifurcated orbits, which 
are transformed into each other by applying $\hat{P}$. Indeed in
Fig.1b we observe two bifurcated orbits. We have tested the permutation
of these orbits numerically, and found that both orbits break the
permutational symmetry and are transformed into each other by applying 
$\hat{P}$. Moreover, all quasiperiodic motions surrounding any one
of the bifurcated orbits and bounded by the separatrix also have
broken permutational symmetry. The whole corresponding phase space
regions are connected to each other by $\hat{P}$.

It is straightforward to show that also the tangent bifurcation
of orbit II creates bifurcated orbits which break the permutational
symmetry. The scenaria of both bifurcations are identical.
This symmetry breaking of the dynamics of two coupled oscillators
has been observed also numerically by Vakakis and Rand \cite{vr92}.

Let us summarize what we have shown so far. We studied the
problem of two coupled oscillators and derived the dependence
of the energy of tangent bifurcation of normal modes 
on the model parameters for
small energies. We have shown analytically and demonstrated
numerically that the new bifurcated orbits will break the permutational
symmetry of the system (if it existed).

\section{The case of the lattice}

Let us now turn to the main subject - the bifurcational
problem of periodic orbits of a lattice. We will consider
a one-dimensional lattice with nearest neighbour interaction
and one degree of freedom per unit cell. We will then show that
the derived results apply also to lattices 
with larger interaction ranges. The problem of lattices
with higher dimension will be discussed.

The Hamilton function of the lattice is given by
\begin{equation}
H = \sum_{l=1}^{N}\left[ \frac{1}{2}\dot{X}_l^2 + V(X_l)
+ \Phi(X_l - X_{l-1})\right]\;\;. \label{20}      
\end{equation}
The potential functions $V(z)$ and $\Phi(z)$ are again represented
in the form (\ref{2-1}),(\ref{2-2}).
We assume periodic boundary conditions
\[
X_{l}=X_{l+N}\;\;,\;\;\dot{X}_l=\dot{X}_{l+N}\;\;.
\]
Then system (\ref{20}) exhibits permutational symmetry. The
permutational operator $\hat{P}$ is defined by
\begin{equation}
\hat{P}g(X_1,X_2,...,X_N,\dot{X}_1,\dot{X}_2,...,\dot{X}_N) =
g(X_2,X_3,...,X_N,X_1,\dot{X}_2,\dot{X}_3,...,\dot{X}_N,\dot{X}_1)\;\;.
\label{21}    
\end{equation}
Clearly $\hat{P}^N=\hat{1}$ and $\hat{P}H=H$.

Let us introduce normal coordinates
\begin{equation}
Q_q=\frac{1}{N}\sum_{l=1}^{N} {\rm e}^{iql}X_l \;\;. \label{22}       
\end{equation}
The wave number $q$ can take any of the values
\[
q=\frac{2\pi}{N}n\;\;,\;\;n=0,1,2,...,(N-1)\;\;.
\]
The inverse transform of (\ref{22}) is given by
\begin{equation}
X_l = \sum_q {\rm e}^{-iql}Q_q \;\;. \label{23}    
\end{equation}
The equations of motion for the normal coordinates $Q_q$ read
\begin{equation}
\ddot{Q}_q = \frac{1}{N}\sum_{l=1}^N {\rm e}^{iql}\ddot{X}_l =
-\frac{1}{N}\sum_{l=1}^N {\rm e}^{iql}\frac{\partial H}{\partial X_l}
\;\;.\label{24}     
\end{equation}
Using (\ref{20}) and (\ref{2-1}),(\ref{2-2}) we obtain the following
lengthy expression
\begin{eqnarray}
\ddot{Q}_q=-\omega_q^2Q_q - \sum_{\mu=3}^{\infty}v_{\mu}
\sum_{q_1,q_2,...,q_{\mu-2}}\left[ \prod_{\nu=1}^{\mu-2}Q_{q_{\nu}}
\right] Q_{q-\sum_{\nu=1}^{\mu-2}q_{\nu}} -\;\;\;\;\;\;\;\;\;\;
\;\;\;\;\;\;\;\;\;\;\hspace*{2cm} \nonumber \\
-\sum_{l=1}^N {\rm e}^{iql} \sum_{\mu=3}^{\infty}\phi_{\mu}
\left[ \left\{ \sum_{q'}(1-{\rm e}^{iq'}){\rm e}^{-iq'l}Q_{q'}\right\}^{\mu-1}
- \left\{ \sum_{q"}({\rm e}^{-iq'}-1){\rm e}^{-iq"l}Q_{q"}\right\}^{\mu-1}   
\right]\;\;.\;\; \label{25}
\end{eqnarray}
Here $\omega_q$ abbrevates the eigenfrequencies of the linearized (in $Q_q$)
equations of motion and is given by the dispersion relation
\begin{equation}
\omega_q^2 = v_2 +4\phi_2 {\rm sin}^2\left( \frac{q}{2}\right)\;\;.
\label{26}       
\end{equation}
Let us give the solutions for two periodic orbits of the considered
lattice, which correspond to the band edge plane waves ($q=0$ and
$q_{N/2}=\pi$) in the limit of small energies:
\begin{eqnarray}
{\rm I:} \;\;Q_{q\neq 0}=0\;\;,\;\;\ddot{Q}_{q=0}=-\sum_{\mu=2}^{\infty}
v_{\mu}Q_{q=0}^{\mu-1}\;\;, \label{27-1} \\
{\rm II:}\;\;Q_{q \neq \pi}=0\;\;,\;\;\ddot{Q}_{q_{N/2}}=
-\sum_{\mu=2,4,6,...}^{\infty}\bar{v}_{\mu}Q_{q_{N/2}}^{\mu-1}\;\;.
\label{27-2}     
\end{eqnarray}
The paramter $\bar{v}_{\mu}$ is given by
\[
\bar{v}_{\mu} = v_{\mu} + 2^{\mu}\phi_{\mu}
\]
(note the difference to (\ref{5-3}). In case II we have to demand
(as in the previous chapters for the two oscillators)
\[
{\rm II:}\;\; v_{2m+1}=0
\]
in order to be able to continue the upper band edge plane wave
to finite energies in the given form. 
The terms $\phi_{2m+1}$ can be in general nonzero,
but they simply do not contribute to (\ref{27-2}) because of the
odd symmetry of the upper band edge plane wave \cite{sp94}.

\subsection{Tangent bifurcation of orbit I}

Let us consider a small perturbation $\left\{ \delta_{q}\right\}$
of the periodic orbit I
\[
Q_q \rightarrow Q_q + \delta_{q}\;\;.
\]
Linearizing the equations of motion for the perturbation we obtain
\begin{equation}
\ddot{\delta}_q = -\omega_q^2\delta_q - \sum_{\mu=3}^{\infty}
(\mu-1)v_{\mu}Q_{q=0}^{\mu-2}\delta_q \;\;.\label{28}       
\end{equation}
The analogy to the problem of two coupled oscillators is striking
(cf. (\ref{5-1}),(\ref{8-1}),(\ref{8-2})). For $q=0$ equation (\ref{28})
describes the continuation of the periodic orbit itself. All other
perturbations do not couple with each other, so that we can consider
(\ref{28}) for each value of $q$ separately. If we increase the
energy 
\begin{equation}
E_I = \frac{1}{2}\dot{Q}_{q=0}^2 + V(Q_{q=0}) \label{29}    
\end{equation}
(note that the periodic orbit solution has to be inserted in (\ref{29}))
then the first tangent bifurcation will occur if $q_1=2\pi /N$
is choosen in equation (\ref{28}). For large values of $N$ we
have 
\[
\omega_{q_1}^2=v_2 +4\phi_2\frac{\pi^2}{N^2}
\]
and consequently obtain (using (\ref{6-1})-(\ref{6-4}) and 
(\ref{9-1})-(\ref{13-3})) the following bifurcation energy $E_I^c$:
\begin{eqnarray}
{\rm i):} \;\; \phi_2=0\;\;,\;\;E_I^c\;\; {\rm arbitrary}\;\;, \label{30-1} \\
{\rm ii):}\;\; E_I^c = \frac{1}{N^2}\frac{12\pi^2 v_2^2 \phi_2}
{10v_3^2 - 9v_2v_4}\;\;. \label{30-2}
\end{eqnarray}
Since we have to require positive values for $E_I^c$ and $\phi_2$
tangent bifurcation can take place only if
\begin{equation}
\frac{v_4}{v_2} \leq \frac{10}{9}\frac{v_3^2}{v_2^2} \label{31}     
\end{equation}
in full accordance to equation (\ref{15}). Note that this condition
is only neccessary, since we have also to require $v_2 \neq 0$ (in
the case $v_2=0$ the dependence of the periodic orbit solution on $E_I^c$
becomes different; since this is a nongeneric case, we ignore it).
Condition (\ref{31}) is equivalent to the condition, that the frequency
of the lower band edge plane wave decreases with increasing energy,
In other words, the necessary condition for a tangent bifurcation
of the lower band edge plane wave is the repelling of its frequency
from the linear spectrum (\ref{26}) with increasing energy.

The energy $E_I$ is on the scale of total energy per particle (cf.
(\ref{20}),(\ref{22}),(\ref{29})). Consequently the
energy threshold of
the tangent bifurcation decreases as $N^{-2}$ on the scale of
energy per particle. In the limit $N \rightarrow \infty$ the threshold
goes to zero. 

It is important to note, that if all variables $X_l$ are real,
the normal coordinates as defined are complex. However since the
equation (\ref{28}) is linear in $\delta_q$ and since the mode
$Q_{q=0}$ is real, we can consider the perturbational problem
always separately for the real and imaginary parts of a perturbation
$\delta_q$.

\subsection{Tangent bifurcation for orbit II}

We again derive the linearized equations for the perturbation $\delta_q$:
\begin{eqnarray}
\ddot{\delta}_{q_{a,b}}=-\omega_{q_{a,b}}^2 \delta_{q_{a,b}}
\sum_{\mu=4,6,...}^{\infty}(\mu-1)\bar{\bar{v}}_{\mu,q_{a,b}}Q_{q_{N/2}}
^{\mu-2} \delta_{q_{a,b}} \nonumber \\ 
-i\sum_{\mu=3,5,...}^{\infty}(\mu-1)\bar{\bar{\phi}}_{\mu,q_{b,a}}
Q_{q_{N/2}}^{\mu-2}\delta_{q_{b,a}}\;\;. \label{32}      
\end{eqnarray}
Here we have used the following notations:
\begin{eqnarray}
\bar{\bar{v}}_{\mu,q_{a,b}} = v_{\mu} + {\rm sin}^2(\frac{q_{a,b}}{2})
2^{\mu}\phi_{\mu} \;\;, \label{33-1} \\
\bar{\bar{\phi}}_{\mu,q_{a,b}}=-{\rm sin}(q_{a,b})2^{\mu}\phi_{\mu}
\;\;. \label{33-2}  
\end{eqnarray}
The two wave numbers $q_{a,b}$ are related to each other by 
\begin{equation}
q_b=q_a\pm \pi \;\; {\rm mod}2\pi\;\;.\label{34}     
\end{equation}
In contrast to the previous cases we have a coupling between a pair
of normal coordinates (\ref{34}) in (\ref{32}). Notice 
that the coupling term is given by the second sum on the right hand
side of (\ref{32}) and is zero if $\phi_{2m+1}=0$. Also this coupling
term is proportional to $i$, which causes a mixing of real and imaginary
parts of the perturbations.

Interestingly for the pair $q_a=\pi,q_b=0$ we obtain again no coupling,
because (\ref{33-2}) vanishes for both wavenumbers. Consequently
 for $q=\pi$
(\ref{32}) describes  the continuation of the periodic orbit II.

\subsubsection{The case $\phi_{2m+1}=0$}

If we assume $\phi_{2m+1}=0$ then the equations for the perturbations
$\delta_{q_{a,b}}$ decouple. In analogy to the case of orbit I
we have the first tangent bifurcation of orbit II if we consider
the perturbation $q=\pi(1-2/N)$. Consequently we obtain for the 
bifurcation energy $E_{II}^c$
\begin{eqnarray}
{\rm i):} \;\;\phi_2=0\;\;,\;\;E_{II}^c\;\; {\rm arbitrary}\;\;, 
\label{35-1} \\
{\rm ii):}\;\; E_{II}^c=\frac{1}{N^2}\frac{v_2+4\phi_2}{3(v_4+16\phi_4)}
4\pi^2\phi_2\;\;.\label{35-2}     
\end{eqnarray}
Again we observe that a bifurcation will take place only if $\bar{v}_{\mu}$
is positive, i.e. only if the frequency of the periodic orbit II is
repelled from the linear spectrum with increasing energy.

We can compare these findings with the calculation of Budinsky and
Bountis \cite{bb83}, who have calculated the bifurcation energy
for $v_{\mu=0}$, $\phi_2=\phi_4=1$, $\phi_{\mu}=0$ otherwise.
They obtain $E^c=3.226/N^2$ (equation (2.22) from \cite{bb83}). Our result
gives $E^c=\pi^2/(3N^2) \approx 3.29/N^2$. 
Budinsky and Bountis have roughly estimated the
eigenvalue spectrum of the corresponding Hill's matrix in the original
coordinates. We think that this circumstance is the reason for
the small deviation of their approximate result from our exact one.
 
Note that Budinsky and Bountis are looking for the first bifurcation
to occur, either tangent or period doubling. It can happen that
a period doupling bifurcation occurs prior the tangent one. In that
case only the result for the period doubling bifurcation would be given
in their paper. In the quoted comparison indeed the tangent bifurcation
takes place first, thus the apparent coincidence in the results. However
for other cases period doubling bifurcations
can occur first. Then our results can not be compared to the ones
of Budinsky and Bountis (cf. also the discussion section).        

\subsubsection{The case $\phi_{2m+1} \neq 0$}

Now the coupling between two perturbations has to be taken into account.
We use the fact that $Q_{q_{N/2}}$ is real and the equations for
the complex perturbations are linear. By taking the complex conjugate
of any of the two equations defined by (\ref{32}) and considering
say only the real part of $\delta_{q_{a}}$ and the imaginary part of 
$\delta_{q_{b}}$ we find in the limit of small energies $E_{II}$
\begin{equation}
\ddot{\delta}_{q_{a,b}}= -(\omega_{q_a}^2+3\bar{\bar{v}}_{4,q_{a,b}}
Q_{q_{N/2}}^2) \delta_{q_{a,b}} -\bar{\bar{\phi}}_{3,q_{b,a}}
Q_{q_{N/2}}\delta_{q_{b,a}}\;\;. \label{36}    
\end{equation}
Note that we have changed the relation between $q_{a}$ and $q_{b}$ to
\[
q_b=|q_a - \pi| \;\;,\;\;q_a\;\; {\rm mod} \pi/2 \;\;.
\]
The lowest tangent bifurcation energy of orbit II appears for
\[
q_a=\pi(1-\frac{2}{N})\;\;,\;\;q_b=\frac{2\pi}{N}\;\;.
\]
Then it follows
\[
\bar{\bar{\phi}}_{3,q_a}=\bar{\bar{\phi}}_{3,q_b}=-16\phi_3 
\frac{\pi}{N}\;\;.
\]
One could expect that in the limit of large $N$ the coupling between
$q_a$ and $q_b$ vanishes. That is indeed so if we require $v_2 \neq 0$
(linear spectrum is optical-like) but turns out to be wrong for the
case $v_2=0$ (linear spectrum is acoustic-like). The details of
these subtleties are given in Appendix A. Here we proceed to
the final result for the tangent bifurcation energy $E_{II}^c$:
\begin{eqnarray}
{\rm i):}\;\;E_{II}^c=\frac{1}{2\phi_3^2}\left( v_2+4\phi_2\right)
\left(\phi_2+\frac{3}{16}v_2\right) \phi_2\;\;, \label{36-1} \\
{\rm ii):} E_{II}^c=\left\{ 
\begin{array}{lr}
\frac{4\pi^2}{N^2}\frac{v_2+4\phi_2}{3(v_4+16\phi_4)}\phi_2 
& v_2\neq 0 \\
\frac{16\pi^2}{N^2}\frac{\phi_2^3}{3\phi_2(v_4+16\phi_4)-64\phi_3^2}
& v_2=0
\end{array}
\right.
\;\;. \label{36-2}  
\end{eqnarray}
As we can see solution i) (\ref{36-1}) is always positive (since
$\phi_2 > 0$ and $v_2 \geq 0$) but is not dependent on $N$. 
Since we applied perturbation theory, (\ref{36-1}) is correct only
in the limit of small energies. For larger energies corrections apply. 
Solution ii) (\ref{36-2}) is the one which gives
arbitrarily small bifurcation energies for sufficiently large $N$.
Since $\phi_3$ does not enter the energy dependence of periodic orbit II,
we then again obtain as the neccessary condition for the
existence of the bifurcation, 
that the frequency of the upper band edge plane wave
has to be repelled from the linear spectrum with increasing
energy. However for the case
$v_2=0$ a more restrictive condition is obtained by demanding
\begin{equation}
3\phi_2(v_4+16\phi_4 ) \geq 64 \phi_3^2 \;\;. \label{37}        
\end{equation}

Sandusky and Page \cite{sp94} have considered the case
$\phi_{2} \neq 0$, $\phi_{3} \neq 0$,$\phi_{4} \neq 0$,$\phi_{\mu}=0$
otherwise. However in contrast to Budinsky and Bountis no clear
finite size studies are done in \cite{sp94}. Instead these authors
use dimensionless parameters, which however contain the 
amplitude (energy). 
Still we can consider equation (24) from \cite{sp94}
and by performing the limit of small amplitudes we recover our
neccessary condition (\ref{37}).
The figures 8 and 9 from \cite{sp94} are showing a crossover from the
parameter region where (\ref{37}) holds into the region where it is
not valid. From our analysis it follows that this corresponds to a change
from (\ref{36-2}) to (\ref{36-1}) and consequently to a change
from infinitesimally small energy (amplitude) thresholds to
finite thresholds. Indeed finite amplitude thresholds
for the instability have been observed in \cite{sp94} after leaving
the parameter region where (\ref{37}) is valid. 

It is interesting to note that condition (\ref{37}) has been obtained
with the help of multiple scale expansions already in 1972 by Tsurui
\cite{at72} (equation (4.8), $\omega_c^2=4$) 
and more recently by Flytzanis, Pnevmatikos and Remoissenet
\cite{fpr85} (equation (5.15)) for systems with $v_{\mu}=0$.

\subsection{Symmetry breaking}

Using (\ref{21}),(\ref{22}) it follows
\begin{equation}
\hat{P}^n Q_q = {\rm e}^{iqn}Q_q\;\;,\;\; q=\frac{2\pi}{N}m\;\;,\;\;
m=0,1,2,...,(N-1)\;\;.\label{38}       
\end{equation}
First we notice that the permutational operator does not mix the
space of the normal coordinates. In other words, every subspace
$\{\dot{Q}_q,Q_q\}$ is invariant under (\ref{21}). 
However since the phase space is higher dimensional compared to
the case of two coupled oscillators, the new bifurcated orbits
can not be confined to some subspace (in order to use
the argumentation of the two coupled oscillators case).

Let us consider the bifurcation of orbit I.
We will project a new bifurcated orbit loop into the
subspace $\{Q_0,\dot{Q}_0,Q_1,\dot{Q}_1  \}$. We can use 
a reference picture in a three-dimensional space, 
where the Z-axis contains both variables
$Q_0$ and $\dot{Q}_0$. Now the projected loop can cross the Z-axis
(in contrast to the case of two coupled oscillators), because at
the moment of the crossing the original loop can still not intersect
with the subspace $\{Q_0 ,\dot{Q}_0\}$. However we notice that 
every loop (also the projected one) have a certain direction of rotation
of the trajectory. If the loop is invariant under $\hat{P}$, so is
its projection. The direction of rotation can either be conserved 
or change sign under permutation. So if we choose 
$n=N/2$, $n$ even, in (\ref{38}) 
then the direction of rotation is conserved. 
Then we observe, that the constructed operation $\hat{P}^{N/2}$
leaves the periodic orbit I, but switches
the sign of the coordinates of the relevant perturbation $Q_1$.
Applying the permutation to the projected loop is equivalent
to reflecting the loop at the subspace $\{Q_0,\dot{Q}_0 \}$ (the Z-axis
in the reference picture). 
Clearly this reflection switches the direction of rotation.
Consequently our initial assumption was wrong, and  
the bifurcated orbits indeed break
permutational symmetry. 
It is straightforward to show that also the new bifurcated orbits
in the case of a tangent bifurcation of orbit II break permutational
symmetry.

Then in general the new bifurcated orbits will
be not invariant under $\hat{P}$, and thus the bifurcated orbits will
correspond to band edge modes. 
which are spatially modulated
by the relevant perturbation. 
This perturbation can be represented as
a wave with a wavelength equal to the length of the system
(see Fig.3). Note that
the  modulation referrs to the {\sl additive} perturbation, so
the bifurcated orbits do not vanish in real space for certain coordinates
$X_l$.  They are thus formally similar to standing waves, but
also inherently different.
Applying a permutational operation, we will simply shift
the modulated object through the lattice. Consequently we arrive
at the result, that after the tangent bifurcation of a band edge
plane wave (orbit I or II) takes place, at least $N$ 
new periodic orbits bifurcate
from the original one. 

It is much harder to argue about the exponential character of spatial 
localization of the bifurcated orbits. 
For any large but finite $N$ the qualitative form of the bifurcated
orbit as shown in Fig.3 is correct if the amplitude of the perturbation
is small compared to the amplitude of the bifurcating orbit (I or II)
at the bifurcation. The frequency of the orbits I (II) and of the 
newly bifurcated ones are outside the linear spectrum. 
Let us consider the analytical continuation of a newly bifurcated
orbit. Its frequency will follow (with deviations of higher order)
the one of the original periodic orbit (I or II), i.e. 
$\Delta(\omega^2) \sim E/N$. So we can assume that the frequency is further
repelled from the linear spectrum. The amplitude of the original periodic
orbit grows like $(E/N)^{1/2}$. The amplitude of the relevant perturbation
however grows much faster. That is due to the fact, that at the
bifurcation at least $N$ new periodic orbits occur. So in terms of
catastrophe theory we would be confronted with an $A_N$ singularity
\cite{rg81}. The
amplitudes of the relevant perturbations will grow as $(E/N-E^c)^{\alpha}$,
where $\alpha \leq 1/2$. Thus
the amplitude of the 
relevant perturbation (describing the deviation of the new periodic
orbits from the original one) will grow much faster than the 
amplitude of the original orbit as a function of energy. Consequently we
can expect, that the overall amplitudes in region II in Fig.3.
will be drastically lowered, whereas in regions I and III they will
grow with energy. This gives us a tendency towards localization, but not
more than that. If it indeed localizes more with increasing energy
then the spatial decay will be exponential in region II, because
the frequency of the bifurcated orbit is outside the linear spectrum.
Clearly it is necessary to explicitely take into account all nonlinearities
in order to give a more precise answer on the question of localization.

\section{Discussion of the results}

We have studied the tangent bifurcation properties of
band edge plane waves in the limit of large system size.
We have restricted the consideration to one-dimensional lattices
with nearest neighbour interaction and one degree of freedom
per unit cell. Clearly the results can be generalized.

First we consider the case of higher dimensional lattices. 
Breathers have been numerically observed in
\cite{fw8},\cite{bkp90},\cite{ff93} and analytically obtained in
\cite{ma94} for selected systems.
Again
the used approach can be applied. It is essential to know the
properties
of the linear spectrum. 
Then one has to evaluate the system size dependence of the 
relevant perturbations (the ones with the eigenvalues closest
to the band edge frequency) and to repeat the calculations.
Without performing this calculation
(which is anyway easy to do using the results of this work)
we mention that the system size $N$ of the one-dimensional lattice
(i.e. the number of degrees of freedom) has to be changed to
$N^{1/d}$ in the expressions for the bifurcation energy
derived in this paper, where $d$ is the dimensionality of the lattice.
So except for some changes in the formula no qualitative differences
appear. Still there is a problem - as one can see, the total
bifurcation energy is independent of $N$ for $d=2$ and even increasing
with $N^{1/3}$ for $d=3$. Using \cite{ma94} we know that in
weakly coupled anharmonic oscillator arrays discrete breathers
exist at arbitrary small energy for zero coupling, and thus at small
energy for weak coupling. Consequently the bifurcation of the 
band edge plane wave for $d=2,3$ might be not 
directly related to these solutions.

We can only suspect, that other channels of breather birth through
bifurcations exist. In any of the considered systems 
standing waves occur in the linearized case which correspond to
resonant tori (two identical frequencies). These standing waves
already in the linear case break the permutational symmetry. 
The interesting cases correspond to wavenumbers closest to
the band edge plane wave. The standing wave would then have two
distinct knots along either spatial direction, since the original
waves had a wavelength equal to the system size.
These resonant tori
will most probably transform into chains of regular islands separated
by a separatrix, once the nonlinearities are included. The newly 
bifurcated orbits can shift their frequencies out of the linear
spectrum with increase of energy. Since they were close to the
band edge plane wave it is likely, that the energy dependence
of the frequencies of these new orbits is similar to the one
of the band edge plane wave. If this scenario is true, then
one could still predict the existence of breathers in higher dimensional
lattices by analyzing the energy dependence of the band edge
plane wave frequency. A logical question would be, whether the same
effect of breather birth through resonant tori (rather then through
tangent bifurcation of a band edge plane wave) takes place
in the $d=1$ systems, which were considered in the present paper.
At the present the answer is, that it might well be. Still there
exist breathers at infinitely small energies which are bifurcating
from the band edge plane wave, as shown in this work, and as
also observed in numerical studies \cite{sp94}. One could argue that
for the $N=40$ example in Fig.8 of \cite{sp94} the dissapearance of
the discrete breather at the frequency of the band edge plane wave
(of same energy) could be nearly reproduced by the breather merging
with a standing wave which has a difference of about $1/1600$ in frequency
compared to the band edge one. But the main problem with the standing
wave is, that the bifurcating objects will have the symmetry of
two breathers. At the present we do not see another channel for
a single breather birth via bifurcation rather than the band edge
plane wave bifurcation discussed in the present paper.              

Secondly we consider the problem of larger interaction range.
Again changes in the linear spectrum have to be considered, and
for the upper band edge plane wave (orbit II) also some changes
in the derivation have to be expected. For 
the case of weak next-to nearest neighbour interaction (compared
to the nearest neighbour one) a smooth change of the derived
results will apply. Especially we observe, that the tendency
is to increase the distances between the eigenvalues of the
linearized spectrum (at fixed $N$) and thus to increase
the bifurcation energy. Thus increasing the interaction
range corresponds to a gradual suppression of bifurcation and localization.
That seems to be plausible, since in certain limits of mean-field
type interaction no localization should occur.

The analysis of systems with several degrees of freedom per unit cell
seems to be more complicated, allthough the presented
approach can be used. It becomes a matter of right dealing with
polarization vectors etc. Here we can only expect that 
tangent bifurcations of band edge modes will appear if
the neccessary condition of the mode frequency repelling from the
linear spectrum (band) occur. 

In this work we have studied only tangent bifurcations of
band edge plane waves. These cases have to be clearly separated
from e.g. period doubling bifurcations (cf. \cite{sp94}). Especially
the subtle connection between tangent bifurcation of band
edge plane waves and the appearance of new periodic orbits
which break permutational symmetry and exponentially localize
in the configurational space is not easily transferred to the 
case of period doubling bifurcations. Since period doubling bifurcations
will occur only for the upper band edge plane wave, the bifurcated
orbits will have frequencies located inside the linear spectrum. Thus
we doubt that any tendency for exponential localization can occur
in these cases.

Let us estimate the bifurcation energy of a period doubling
bifurcation of periodic orbit II. The relevant perturbation
is given by $q=\pi/6$ for the one-dimensional case
(the upper band edge of the linear spectrum is given
by $2\sqrt{\phi_2}$, the relevant frequency of the perturbation
has to be close to $\sqrt{\phi_2}$). If $N/6$
is an integer, then we have to expect a period doubling bifurcation
at zero energies, since the normal modes resonate already in the linear
problem. If $N/6$ is noninteger, then the distance of the closest
eigenvalue of the linear spectrum to the value $\phi_2$ is of the
order $1/N$. Note that this is different from the tangent bifurcation
cases considered above, where the distance of the closest eigenvalue
to the squared band edge was always of the order $1/N^2$. Thus
we conclude that for the generic cases period doubling bifurcations
occur at energies $E^c \sim 1/N$
(cf. Budinsky and Bountis \cite{bb83}, equation (3.8))
 whereas the tangent bifurcation
occurs at energies $E^c \sim 1/N^2$.
So generically tangent bifurcations appear at lower energies than
period doubling bifurcations. Then we can conclude, that with
increase of energy first exponentially localized orbits (discrete
breathers) are generated through tangent bifurcations. 

If we consider a nonlinear Klein-Gordon lattice (optical-like
linear spectrum, $v_2 \neq 0$) then for small bandwidth 
($\phi_2 \ll v_2$) no period-doubling bifurcations occur at small
energies. Since generically any nonlinearity will repell
the frequency of one of the two band edge plane waves with
increase of energy, tangent bifurcation always occurs. This
result is in agreement with MacKay and Aubry \cite{ma94}, 
who have shown that
discrete breathers exist for these lattices for any nonlinearity,
provided the bandwidth is small.

Let us give an example where no discrete breathers are expected.
The one-dimensional Toda chain is characterized by \cite{mt89}
$v_{\mu}=0$, $\phi_2=1$, $\phi_3=-1/2$, $\phi_4=1/6$. Clearly (\ref{37})
is not satisfied (note that in that case the frequency of the
upper band edge plane wave {\sl is} repelled from the linear 
spectrum with increase of energy, so the repelling condition is 
a necessary condition, but not a sufficient one).
Thus in the limit of large system size 
no tangent bifurcation occurs at small energies, 
and no discrete breathers should 
occur. This circumstance should make clear, that discussions
about 'nonexistence of breathers in systems with realistic potentials'
are simply based on the fact, that a potential called realistic
can be choosen in a way that (\ref{37}) does not allow for tangent
bifurcation of an upper acoustic band edge phonon. 

A few words should be added in order to explain the subtleties
of the result of the previous paragraph. If we consider a system
with an acoustic-type linear spectrum (like that of the Toda chain)
then we deal with equations (\ref{36-1}) and (\ref{36-2}). 
The first tangent bifurcation occurs at the value of energy which
corresponds to the smallest nonnegative outcome of any of both
equations.
If (\ref{37})
is fulfilled, then (\ref{36-2}) gives positive energies, which become
arbitrarily small as the size of the system is increased to infinity. 
Consequently
the new bifurcating orbits have zero amplitudes and energy densities
outside their modulation center and eventually evolve into
discrete breathers with increase in energy. If however (\ref{37}) does not hold
(e.g. as in the case of the Toda chain) then (\ref{36-2}) gives negative
bifurcation energies. So we have to use (\ref{36-1}), which gives {\sl
finite} energy values even for infinite system size. Despite the fact
that (\ref{36-1}) is only a result of perturbation theory, and
corrections apply for finite but not too small energies, we can conclude
for sure that the tangent bifurcation energy in this case stays finite
with increase of the size of the system. So the new bifurcated orbits
will have {\sl finite nonzero} amplitudes and energy densities outside
their modulation center even if being infinitely close to the bifurcation.
Those objects can not be called discrete breathers, because their amplitudes
can not decay exponentially in space, and because their total energy 
on an infinite lattice will be infinite. Rather they correspond to
 discrete breathers 'sitting' on a {\sl nondecaying} carrier wave.
We suspect that precisely those solutions were observed in Figs.8,9 in
\cite{sp94} but were not identified as being different from discrete breathers
with exponential spatial decay. So the result of the previous paragraph
is that there exist no discrete breathers in a Toda chain. But the next
result and {\sl prediction} is that there exist discrete breathers on
carrier waves in the Toda chain (this last statement should be still
considered to have some uncertainty, because of the mentioned use of
perturbation theory for deriving ({\ref{36-1}})).

Nonlinearity shifts frequencies of periodic orbits with
change of energy. For an optical band no matter whether the
shift results in an increase or decrease of the frequency, 
tangent bifurcation will take place nearly always - either
for the upper or for the lower band edge phonon. In the
case of an acoustic band we need increase of frequencies in order
to obtain tangent bifurcation, and even this condition is only
a necessary one (cf. the two previous paragraphs). 
Thus acoustic band related breathers will occur rarely compared
to the optical band related ones.
This circumstance is in accordance with numerical findings \cite{kbs93},
\cite{ats93}
where periodic modulation of masses (diatomic chain instead of
monoatomic chain) allows for tangent bifurcation of band edge
plane waves of the optical band (which is absent in the monoatomic case,
as were the breathers for the choosen parameters).

From our results we can follow, that breathers will be likely to occur
in lattices with optical bands in the linear spectrum, which are
narrow and well separated from the rest of the spectrum
(we have recently learned about a numerical study of Bonart,
Meyer and Schr\"oder \cite{bms95}, where similar guides have been successfully
applied in order to obtain discrete breathers on surfaces of 3d
crystals). Of course
the nonlinearity has to be strong enough - that can be accomplished
either by not too low temperatures (especially for systems with
structural phase transitions) or by local excitation of the lattice
using e.g. laser pulses, so that sufficiently large amplitudes occur,
which should then provide for strong enough nonlinearities in the
forces. Breathers will certainly influence the thermal conductivity
acting as strong scatterers of small-amplitude phonons \cite{fwdna} and
may also influence other transport properties of other degrees of freedom
(e.g. electrons) if the coupling is strong enough. 

Finally we want to note, that for systems with acoustic-like
linear spectra (say $v_{\mu}=0$) and $\phi_{2m+1}\neq 0$ 
numerical simulations of discrete breathers in one-dimensional
chains always lead
to the presence of static displacements (dc terms) in the
solution $X_l(t)$ which seemingly decay linearly in space,
due to the periodic boundary conditions \cite{sp94},\cite{bks93},
\cite{hsx93}. In the limit of
large system sizes the slope goes as $1/N$ and thus 
vanishes in the limit $N \rightarrow \infty$. This result
can be immediately obtained using the fact that the decay
rate of the dc term is zero for an infinite lattice with
an acoustic linear spectrum \cite{fw7}.
Thus discrete breathers in the mentioned lattice types are
similar to a kink+localized vibration for the one-dimensional
case, and to a kink bubble+localized vibration in higher
dimensional lattices. 
This statement steems from the observation, that far away
from the center of the breather the static displacements
can be described as a strain field caused by a point-like 
source (in full analogy to the corresponding
Maxwell equation for the electric field) \cite{LLVII}. Then the strain field
will be radial and isotropic, and its decay can be calculated
using Gauss's theorem. The resulting decay is $\sim 1/r^{d-1}$,
where $d$ is the dimension of the system. So for three-dimensional
lattices we obtain $1/r^2$ decay, for two-dimensional systems
$1/r$ decay and for one-dimensional systems non-decaying strains.
Consequently even in three-dimensional lattices these discrete
breathers (which bifurcated from an acoustic band
in the presence of $\phi_{2m+1}\neq 0$) 
are composed out of a dynamical (time-periodic) solution
decaying exponentially fast, and a strain field decaying as $1/r^2$.
If we choose $v_2 \neq 0$ (i.e. if we consider a
breather which bifurcated from an optical band) then the spatial decay
of the dc terms becomes exponential. Similar to that we
find that the tangent bifurcation energy of the
upper band edge plane wave behaves also differently for 
the two cases (\ref{36-2}). Performing first the limit
$N \rightarrow \infty$ we thus find that the two different
classes of lattices (acoustic-like or optic-like linear spectra)
have no simple contact. This is connected to the circumstance,
that the case of acoustic-like linear spectra implies the
existence of an additional integral of motion - the total
mechanical momentum. In the case of systems with more than one degrees
of freedom per unit cell it can happen, that tangent bifurcation
appears for an optical band edge plane wave. However the nonlinearities
can couple the band variables of the optical band with the acoustic
band variables. Then the existence of a local strain due to the
discrete breather will still cause a nonexponential decay of the
strain as described above. Thus we agree with the conclusion in 
\cite{kbs93} (where a nondecaying strain was found in a one-dimensional
diatomic lattice - with the breather bifurcating from the optical
band) that the existence of breathers in crystals should
be in general accompanied with anomalies in the thermal expansion.

There exist approximate methods of testing whether
a lattice allows for the existence of breathers 
\cite{fw2},\cite{fw8},\cite{fw10}.
These methods use the hypothetical existence of breathers
and construct then energy dependencies of their frequencies.
If the breather frequencies are attracted by the linear spectrum,
then the conclusion was that breathers most probably do not exist.
These predictions were usually successful.
However this method of argumentation is not well-defined, since it
uses objects which are not defined (as long as we do not know
whether breathers exist, we can not calculate any property of a
breather). In the present work instead a well-defined and clear
criterion is given - whether or not the frequency of a band edge
plane wave is repelled from the linear spectrum with increase in energy.
There should be even experimental methods of testing this property
of a lattice, if the first corrections to the linear spectrum
due to nonlinearity can not be well estimated theoretically.

\section{Summary}

We have shown, that band edge plane waves of nonlinear
Hamiltonian lattices undergo tangent bifurcations. A necessary
condition for that is the repelling of the frequency of the
plane wave from the linear spectrum with increase of energy.
If this condition is fulfilled, then the bifurcation energy
(in units of one-particle energy) scales as $N^{-2/d}$ where
$d$ is the dimensionality of the lattice. The new bifurcated
periodic orbits break the permutational symmetry of the
lattice (if periodic boundary conditions are used). The shape
of the new bifurcated orbits corresponds to a localized vibration.
It is argued that the spatial decay of the amplitudes becomes
exponential with further increase of energy. Thus we confirm
the numerical findings of \cite{sp94}, who have shown that
the new bifurcated orbits are related to discrete breathers in finite chains. 
In accord with the strong evidence that discrete breathers
are generic solutions of nonlinear lattices we find that
tangent bifurcation of band edge plane waves is 
generic too.  
\\
\\
\\
Acknowledgements
\\
\\
It is a pleasure to thank H. Kantz, E. Olbrich and K. Kladko 
for stimulating discussions, J. B. Page for helpful comments,
N. Flytzanis for drawing attention to
the results in \cite{at72},\cite{fpr85} and D. Bonart and A. P. Mayer for 
sending their manuscript before publication.

\newpage

\appendix

\section{Perturbation theory for bifurcation of orbit II in the lattice}

Here we want to scetch the derivation of the critical lines
of tangent bifurcation for periodic orbit II (upper band edge
plane wave) for $\phi_{2m+1}\neq 0$. We start from 
the set of two equations (\ref{36})
and rewrite them in the following way:
\begin{eqnarray}
\ddot{x}=-\left[ \Omega_x^2 + 2F_2^{x}{\rm cos}(2\omega t)\right]
x - 2A \alpha {\rm cos}  (\omega t)   y \;\;, \label{b1-1} \\  
\ddot{y}=-\left[ \Omega_y^2 + 2F_2^{y}{\rm cos}(2\omega t)\right]
y - 2A \alpha {\rm cos}  (\omega t)   x \;\;. \label{b1-2}
\end{eqnarray}
Here we used the different notations
\[
x=\delta_{q_a}\;\;,\;\;y=\delta_{q_b}\;\;,\;\;\omega^2_{x,y}=
\omega^2_{q_{a,b}}\;\;,\;\;\alpha = \bar{\bar{\phi}}_{3,q_{a,b}}\;\;.    
\]
The other parameters are defined in the limit of small energies
of the band edge plane wave as
\begin{eqnarray}
\Omega_{x,y}^2=\omega_{x,y}^2 + F_{0}^{x,y}\;\;,\label{b2-1} \\
F_{0}^{x,y}=3 \frac{\bar{\bar{v}}_{4,q_{a,b}}}{\bar{v}_2}E\;\;,\label{b2-2} \\
F_2^{x,y}= \frac{1}{2} F_0^{x,y}\;\;,\label{b2-3} \\
A= a_1E^{1/2}\;\;,\;\;a_1^2=\frac{1}{2\bar{v}_2}\;\;,\label{b2-4} \\
\omega_x^2=v_2 + 4\phi_2 - \frac{4\pi^2}{N^2}\phi_2 \;\;,\label{b2-5} \\
\omega_y^2=v_2 +  \frac{4\pi^2}{N^2}\phi_2 \;\;.\label{b2-6}       
\end{eqnarray}
We expand the solutions of the
differential equations (\ref{b1-1}),(\ref{b1-2}) into series of
powers of $E^{1/2}$:
\begin{eqnarray}
x(t)=x_0(t)+E^{1/2}x_1(t)+Ex_2(t) + ...\;\;,\label{b3-1} \\
y(t)=y_0(t)+E^{1/2}y_1(t)+Ey_2(t) + ...\;\;,\label{b3-2} \\
\Omega_x^2=\omega^2 +g_1 E^{1/2} + g_2E + ... \;\;.\label{b3-3}  
\end{eqnarray}
Note that we did not expand $\Omega_y$, since in the given
problem this frequency is far away from the frequency of the
periodic orbit $\omega$.

Inserting (\ref{b3-1})-(\ref{b3-3}) into (\ref{b1-1}),(\ref{b1-2})
and sorting with respect to powers of $E^{1/2}$ we obtain
in lowest order
\begin{equation}
\ddot{x_0}=-\omega^2x_0\;\;,\;\;\ddot{y_0}=-\omega_y^2y_0\;\;.\label{b4}        
\end{equation}
Since we are looking for solutions periodic with $2\pi/\omega$
it follows
\begin{equation}
x_0=a{\rm cos}(\omega t) + b{\rm sin}(\omega t)\;\;,\;\; y_0=0\;\;.
\label{b5}      
\end{equation}
In order $E^{1/2}$ we get
\begin{equation}
\ddot{x_1}=\omega^2x_1 - g_1 x_0\;\;,\;\;
\ddot{y_1}=-\omega_y^2y_1 -2\alpha a_1{\rm cos}(\omega t) x_0\;\;.
\label{b6}       
\end{equation}
The solution reads
\begin{eqnarray}
g_1=0\;\;,\;\;x_1=0\;\;,\label{b7-1} \\
y_1=\kappa + c{\rm cos}(2\omega t) + d{\rm sin}(2 \omega t)\;\;,
\label{b7-2} \\
\kappa=-\frac{\alpha a_1}{\omega_y^2}a\;\;,\;\;
\frac{c}{a}=\frac{d}{b}=\frac{\alpha a_1}{4\omega^2-\omega_y^2}\;\;.
\label{b7-3}        
\end{eqnarray}
Note that the solution $x_1=0$ is not strictly required at this level.
However we would have to do so at the next level, so we remove it already
here. The constant $\kappa$ is inverse proportional to $\omega_y^2$.
This circumstance will lead to the difference in the final results
for the optical like spectrum ($\omega_y^2$ finite) and the acoustic
like spectrum ($\omega_y^2 \sim 1/N^2$). 

In order $E$ we obtain
\begin{eqnarray}
\ddot{x_2}=-\omega^2x_2-g_2x_0-f_{2,x}{\rm cos}(2\omega t)x_0
-2\alpha a_1 {\rm cos}(\omega t) y_1\;\;,\label{b8-1} \\
\ddot{y}_2=-\omega_2^2 y_2\;\;.\label{b8-2}          
\end{eqnarray}
Equation (\ref{b8-2}) simply requires $y_2=0$. Removing
the secular terms from equation (\ref{b8-1}) it follows
\begin{eqnarray}
{\rm  (i):}\;\;E^c=\frac{4\pi^2}{N^2}\phi_2 
\frac{4\omega^2-\omega_y^2}{a_1^2 \alpha^2}\;\;,\label{b9-1} \\
{\rm ii):}\;\;E^c=\frac{4\pi^2}{N^2}\phi_2 
\frac{1}{2f_{2,x}-\left(\frac{2a_1^2}{\omega_y^2}-\frac{a-1^2}{4\omega^2
-\omega_y^2}\right)\alpha^2}\;\;.\label{b9-2}           
\end{eqnarray}
Note that $\alpha^2\sim 1/N^2$. Thus we obtain that the $N$-dependence
is removed from equation (\ref{b9-1}).
Now we see the cause for the mentioned difference between the optical
and acoustic spectra. If $\omega_y^2$ stays finite for large $N$,
then the whole second term in the denominator of the right hand side
of equation (\ref{b9-2}) scales to zero. In the case of the
acoustic spectrum $\omega_y^2 \sim 1/N^2$, and the same term in (\ref{b9-2})
gives a finite contribution in the limit of large $N$.

\newpage

FIGURE CAPTIONS
\\
\\
\\
Fig.1
\\
\\
Poincare plot for the system of two coupled oscillators.
The condition is $X_2=0, \dot{X}_2 > 0$. The parameters
are given in the text. Only a part of the available phase
space is shown, which includes the relevant periodic orbit I.
\\
a) The energy is below the bifurcation;
\\
b) The energy is above the bifurcation. A and B labels the
two new periodic orbits, which appeared due to the bifurcation
of the periodic orbit I.
\\
\\
\\
Fig.2
\\
\\
The imaginary part of the Floquet multiplier (absolute value) of
the Hill's equation as a function of the energy $E_2$ 
for periodic orbit II and parameters as given
in the text. At energies larger than the bifurcation energy
the imaginary part vanishes.  
\\
\\
\\
Fig.3
\\
\\
Schematic representation of the relevant perturbation of
a periodic orbit I due to the symmetry breaking. The x-axis
represents the normalized spatial variable $l/N$, and the y-axis
the amplitude of the solution in arbitrary units. The dashed line
indicates the amplitude distribution of the original periodic orbit I.

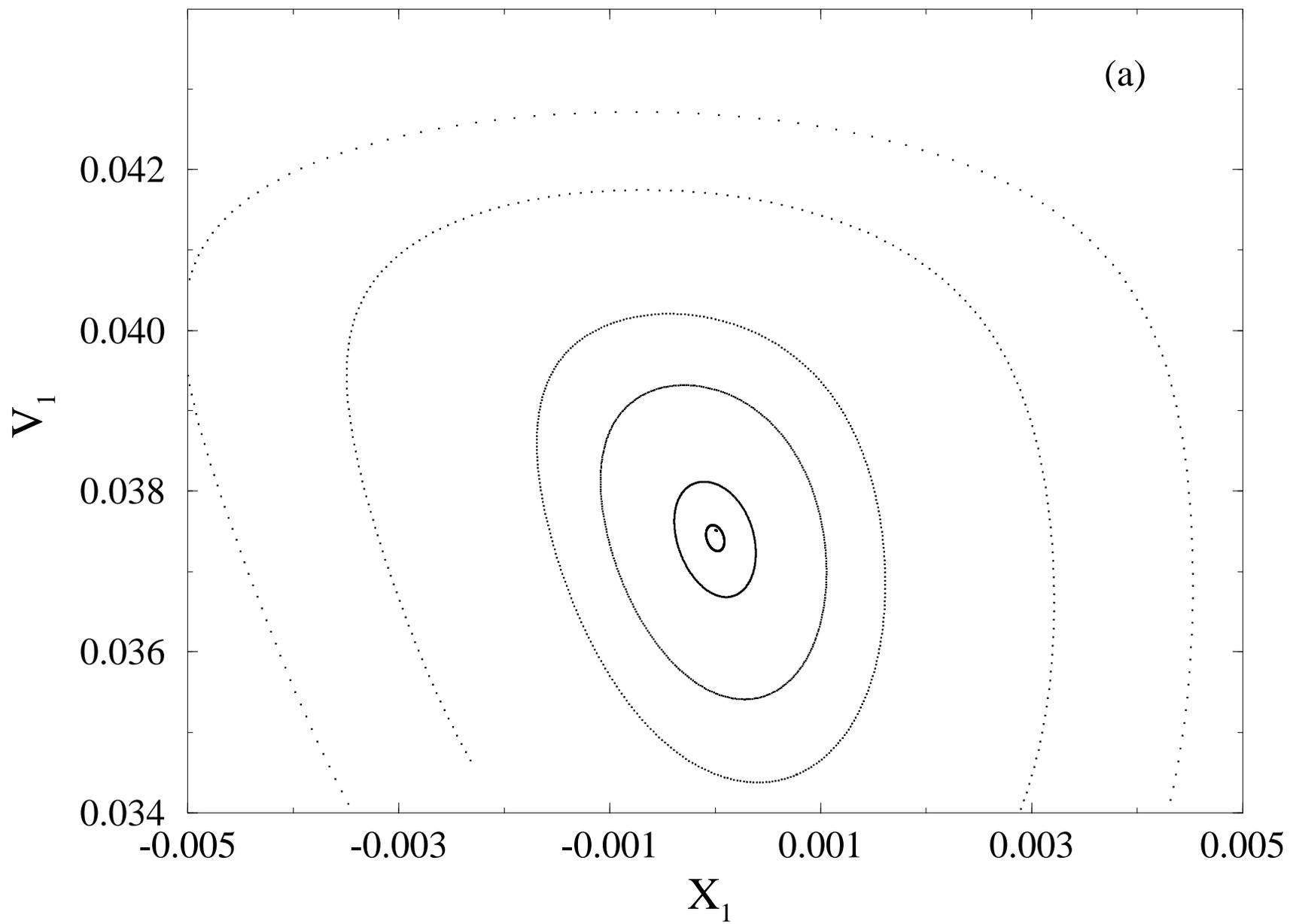
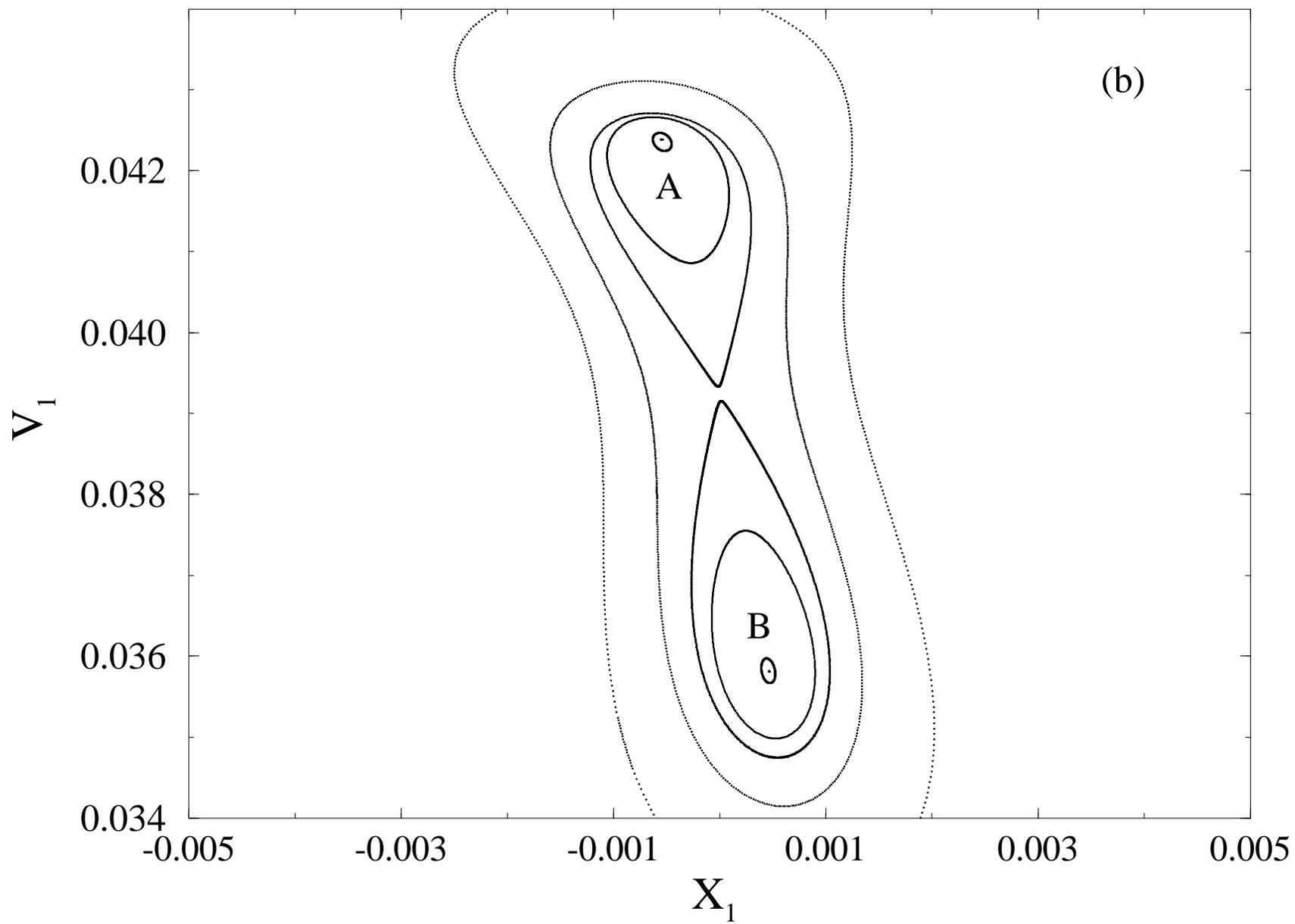
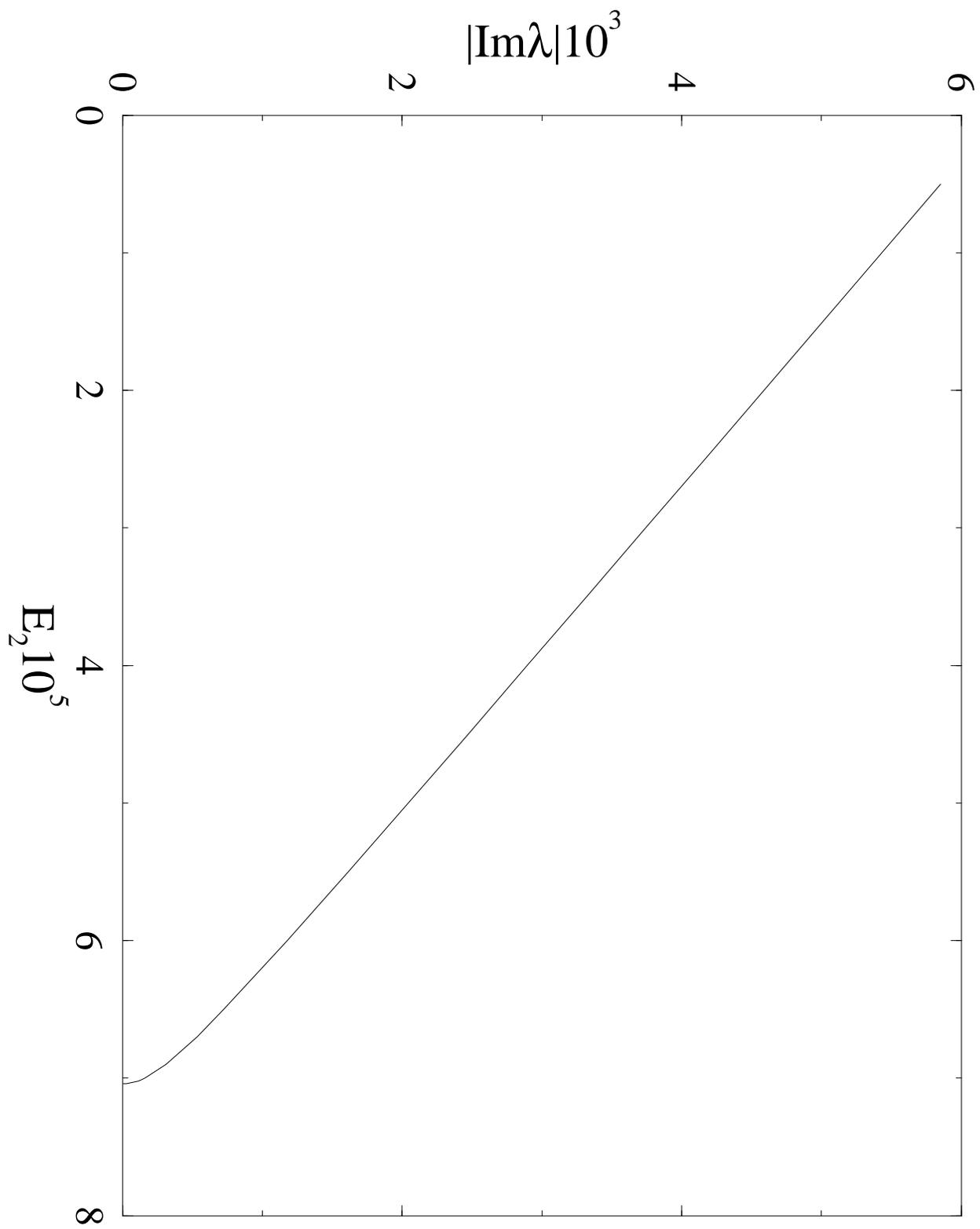
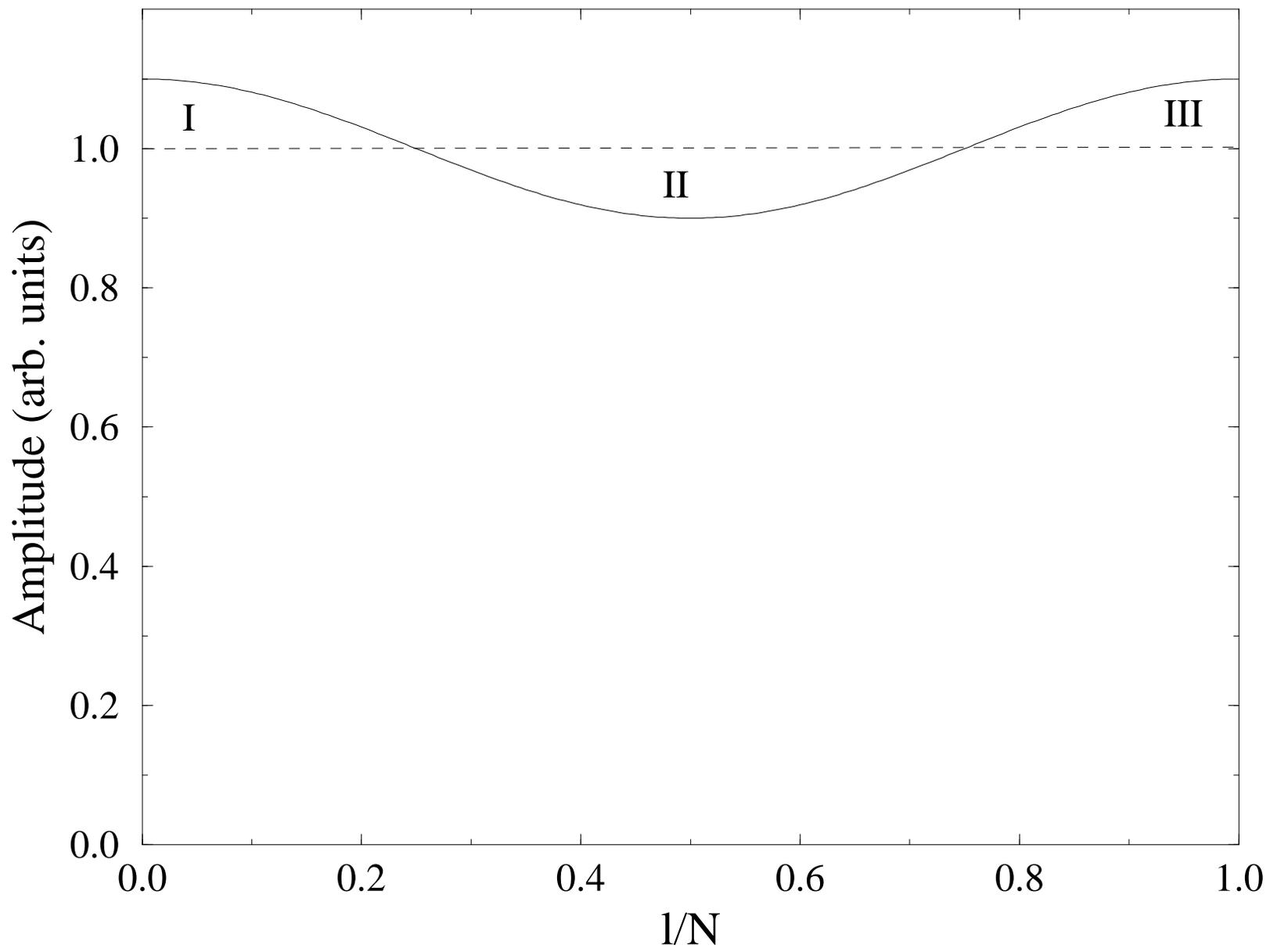
\end{document}